\documentclass[12pt,a4paper]{article}
\usepackage[onehalfspacing]{setspace}

\input{tcilatex}
\begin{document}

\title{Repeat Voting:\\
Two-Vote May Lead More People To Vote\thanks{%
The idea of repeat voting evolved following discussions at the Annual
Conference of the Federmann Center for the Study of Rationality in February
2017. The author thanks the Center's members, in particular Maya Bar-Hillel,
Orit Kedar, and Motty Perry, for their suggestions.}}
\author{Sergiu Hart\thanks{%
The Hebrew University of Jerusalem (Federmann Center for the Study of
Rationality, Department of Economics, and Institute of Mathematics). \emph{%
E-mail}: \texttt{hart@huji.ac.il} \emph{Web site}: \texttt{%
http://www.ma.huji.ac.il/hart}}}
\date{October 17, 2017}
\maketitle

\begin{abstract}
A \textbf{\emph{repeat voting}}\textsc{\ }procedure is proposed, whereby
voting is carried out in two identical rounds. Every voter can vote in each
round, the results of the first round are made public before the second
round, and the final result is determined by adding up all the votes in both
rounds. It is argued that this simple modification of election procedures
may well increase voter participation and result in more accurate and
representative outcomes.
\end{abstract}

Suppose that it is two weeks after the Brexit vote, and there is a new vote
on the same issue---what will the result be? Given the way the original vote
went, will people change their minds and vote differently? Will the original
results cause people who had not voted to cast their vote in this second
round? Will the final result be different?\footnote{%
Before the Brexit vote, a petition that called for a second vote in case of
low participation and a narrow winning margin was launched; it got 22
signatures before the vote, and more than 2 million signatures in the two
days after the result was announced. (Interestingly, the initiator was a
\textquotedblleft leave" supporter who believed that \textquotedblleft
leave" would lose.) See, e.g., \texttt{%
http://www.bbc.com/news/uk-politics-eu-referendum-36629324}} (There are no
clear answers to any of these questions, but one can easily provide
arguments either way.) Now carry out a similar thought experiment regarding
the latest presidential election in the U.S., or whatever your latest
favorite, or unsettling, election is ...

Democratic elections are beset by many problems. One issue is low voter
turnout, which at times is only one-half of the eligible voters or even
less. Another issue is excessive reliance on polls: polls affect voters,
despite repeatedly turning out to being quite far from accurate. This also
relates to the low-turnout issue: \textquotedblleft I will not waste my time
voting, as my candidate is in any case sure to win" (or \textquotedblleft
... sure to lose\textquotedblright ).\ Polls may also lead people not to
cast their vote for their preferred candidate, if, for example, they do not
want him or her to win by too large a majority, or if they want to voice a
certain \textquotedblleft protest" through their vote---only to find out
that in the end their candidate did not win at all. Yet another issue
concerns unexpected events that occur extremely close to election time, too
late to be able to be addressed by the candidates, such as a terrorist
attack, the publication of false information, bad weather, and so on. What
is common to many of these situations is that people might want to change
their vote, or their non-participation in the election, once they see the
actual results and how these came about.

To address these and other issues, I\ propose the use of the following 
\textsc{REPEAT VOTING}\textbf{\ }procedure.%
%TCIMACRO{%
%\TeXButton{TeX field}{\renewcommand{\theenumi}{\textbf{\Alph{enumi}}}}}%
%BeginExpansion
\renewcommand{\theenumi}{\textbf{\Alph{enumi}}}%
%EndExpansion

\begin{enumerate}
\item \textbf{Voting is carried out in two rounds.}

\item \textbf{Every eligible voter is entitled (and encouraged) to vote in
each of the two rounds.}

\item \textbf{All the votes of the two rounds are added up, and the final
election result is obtained by applying the current election rules\footnote{%
Be they plurality, special majority, electoral college, and so on.} to these
two-round totals.}

\item \textbf{The results of the first round are officially counted and
published; the second round takes place, say, two weeks after the first
round, but no less than one week after the official publication of the first
round's results.}
\end{enumerate}

What are the \emph{advantages} of repeat voting?%
%TCIMACRO{\TeXButton{TeX field}{\renewcommand{\theenumi}{\arabic{enumi}}}}%
%BeginExpansion
\renewcommand{\theenumi}{\arabic{enumi}}%
%EndExpansion

\begin{enumerate}
\item \emph{Polls. }The first round becomes a de facto giant opinion poll;
however, because the votes of the first round count, it is a much more
truthful poll (in contrast to the usual pre-election polls, where giving
untruthful answers---whether intentionally or not---carries no cost\footnote{%
Someone once quipped that Israelis tell the truth in polls, but lie when
they cast their vote.}). The combination of the large sample size and
incentivized truthfulness makes the results of the first round a
significantly more accurate predictor of the electorate views. It is thus
crucial for the votes of the first round to count no less than the votes of
the second round,\ which explains why we are adding up the votes of the two
rounds, rather than having only the second round determine the outcome.

\item \emph{Participation. }Voters who do have a preference that is not
however strong enough to make them vote in the first round may well be led
to vote in the second round because of the results of the first round. Thus,
participation in at least one round of the election is expected to increase.
It is better that people vote even in one round than not at all.\footnote{%
Voters who have strong or extreme positions will most probably vote in both
rounds; their relative weight in the final result will decrease when enough
people are motivated to vote in the second round\ (which may well happen if
such extreme positions get higher shares of the vote in the first round).}
One indirect advantage is that people who vote may feel closer to the
elected officials, and to the democratic system in general.

\item \emph{Representative results. }The final results may be more
representative, because the second round makes it possible for the voters as
well as for the candidates to \textquotedblleft correct" any problems of the
first round. This includes the effects of wrong predictions by the polls, as
well as any special circumstances and events that occurred close to election
time (see the second paragraph of the paper; it is unlikely that such
unexpected events will happen both times). All this, again, can only
increase the robustness of the results: they become more trustworthy and
more accepted.

\item \emph{New reference point. }The results of the first round become a
new reference point, which may well affect a person's choice in the second
round: \emph{imagining} a new situation and \emph{being} in a new situation
are not the same thing.\footnote{%
Robert J. Aumann, awarded the Nobel Prize in Economics in 2005, tells the
following story (S. Hart, \textquotedblleft An Interview with Robert
Aumann,\textquotedblright\ \emph{Macroeconomic Dynamics} 9, 2005, page 711;
reprinted in: Paul A. Samuelson and William A. Barnett, editors, \emph{%
Inside the Economist's Mind: Conversations with Eminent Economists},
Blackwell Publishing 2006). In 1956 he had two offers: one from Bell Labs in
New York, and another from the Hebrew University of Jerusalem. It took him a
long time to make up his mind, and he chose Bell Labs. He phoned them and
told them that he accepted their offer. Once he put down the phone, he
immediately started imagining the next few years at Bell Labs, and reached
the conclusion that he had made the wrong choice. A day later he phoned Bell
Labs and asked them if he could change his mind---which they graciously
agreed to. How come a leading game theorist couldn't understand all this
before he made his initial decision? Aumann's answer is that until he found
himself in the new situation of someone going to Bell Labs, he could not
really grasp what it meant!}

\item \emph{Strategic voting. }People seem to be more strategic in their
voting than is usually believed (again, see the examples in the second
paragraph above), but under current procedures they base their strategic
decisions on possibly inaccurate polls. Repeat voting provides a much more
solid basis. In close elections it is conceivable that the voting of the
second round may be less strategic (and the other way around when there is a
large winning margin in the first round).\footnote{%
An interesting related instance concerns the minimal threshold for a party
to be represented in a parliament. Many potential entrants try to convince
voters that they have support that is higher than the threshold and so
voting for them would not be a \textquotedblleft waste" of one's vote. In
many cases, however, it turns out that these parties do not pass the
threshold; once this is seen in the first round, there will be many fewer
such wasted votes in the second round.}
\end{enumerate}

What are the possible \emph{disadvantages} of repeat voting?

\begin{enumerate}
\item \emph{Costs. }A second round adds costs (however, in future voting
that may be conducted online, these costs would become much smaller). The
additional electoral campaign between the two rounds also increases the
costs (but one should remember that two rounds are already used in various
elections, albeit not two identical rounds as proposed here). One way to
save costs is to carry out the second round only when the results of the
first round are close (for instance, when the winning margin is below a
certain threshold that is specified in advance).\footnote{%
Suggested by Motty Perry and Steve Brams.}

\item \emph{Participation. }There may be fewer voters in the first round
(\textquotedblleft I will have a chance to vote in the second round").

\item \emph{Bandwagon effect. }Voters with strong or extreme positions, who
are much more likely to vote in the first round, may have a big effect on
the results of the first round, which may then have a bandwagon effect on
the whole election.
\end{enumerate}

One can think of other ways to overcome the issues pointed out above. For
example, one can repeat the vote three times, with the winner having to win
at least two rounds (this applies only to two-outcome elections, however,
not to multi-candidate and parliamentary elections, and is inherently more
complicated).\footnote{%
This procedure was also suggested by Shachar Kariv.} Another possibility is
to make voting mandatory (as in certain countries); while this may resolve
the participation issue, it does not resolve the significant
\textquotedblleft polls issue" discussed in advantage \#1 above. Yet another
is to have the votes in the two rounds of repeat voting weighted differently
(for instance, depending on the total number of votes in each round\footnote{%
For example, averaging the percentages of votes that each candidate received
in the two rounds (which amounts to giving weights to the two rounds that
are inversely proportional to the total number of votes in each) may perhaps
increase participation in that round where there are fewer voters (probably
the first round).}); at this point, however, it seems best to leave it as
simple and straightforward as possible.

In summary:\textbf{\ }\textsc{REPEAT VOTING }is a simple modification of
election procedures that is capable of increasing voter participation and
yielding more accurate and representative results. Everyone deserves a
second chance, as the saying goes. Shouldn't this include voters and
candidates?

\end{document}